\documentclass[twocolumn,aps,pra,showpacs,amsfonts]{revtex4}

\def\endproof{\vrule height6pt width6pt depth0pt}
\newtheorem{lemma}{Lemma}

\usepackage{graphicx}

\begin{document}
\title{Supersinglets}
\author{Ad\'{a}n Cabello}
\email{adan@us.es}
\affiliation{Departamento de F\'{\i}sica Aplicada II,
Universidad de Sevilla, 41012 Sevilla, Spain}
\date{\today}
\begin{abstract}
Supersinglets $|{\cal S}_N^{(d)}\rangle$ are states of total spin
zero of $N$ particles of $d$ levels. Some applications of the
$|{\cal S}_N^{(N)}\rangle$ and $|{\cal S}_N^{(2)}\rangle$ states
are described. The $|{\cal S}_N^{(N)}\rangle$ states can be used
to solve three problems which have no classical solution: The
``$N$ strangers,'' ``secret sharing,'' and ``liar detection''
problems. The $|{\cal S}_N^{(2)}\rangle$ (with $N$ even) states
can be used to encode qubits in decoherence-free subspaces.
\end{abstract}
\maketitle

\section{Supersinglets}
\label{sec:1}

Quantum information has changed not only the way we understand
quantum mechanics; it has also changed the way we use quantum
mechanics in our dealings with the real world. For years, the
research on quantum mechanics has focused on pointing out how
different quantum mechanics was from classical physics. The next
step has been to realize that quantum mechanics can be used to
solve problems unsolvable by any other means. However, although
quantum states are powerful tools, they are very fragile for most
real-life applications. Therefore, the challenge is twofold: on
one hand, to find out new applications---problems without
classical solutions that can be solved with the aid of quantum
states; on the other, to find out how to protect quantum
states---methods that allow us to use quantum states for practical
purposes. In this paper I review some results regarding a
particular family of quantum states. The properties of these
states, particularly their symmetry, grant them a prominent role
in this challenge.

Supersinglets $|{\cal S}_N^{(d)}\rangle$ are states of total spin
zero of $N$ particles, all of them of spin $(d-1)/2$.
Supersinglets are $N$-lateral rotationally invariant. This means
that, if we act on any of them with the tensor product of $N$ equal
rotation operators, the result will be to reproduce the same state
(within a possible phase factor):
\begin{equation}
R^{\bigotimes N} \left|{\cal S}_N^{(d)}\right\rangle= \left|{\cal
S}_N^{(d)}\right\rangle,
\end{equation}
$R^{\bigotimes N}$ being $R \otimes \ldots \otimes R$,
where $R$ is a rotation operator.

In this paper, $|01\ldots 2\rangle$ denotes the tensor product
state $|0\rangle \otimes |1\rangle \otimes \ldots |2\rangle$,
where $\left| i \right\rangle$ corresponds to the state with
eigenvalue $(d-1)/2-i$ of the spin along the $z$-direction. The
basis we shall use, $B_d=\left\{|i\rangle
\right\}_{i=0}^{d-1}$, and the basis of eigenvectors of the spin
along the $z$-direction, $B_s=\left\{ |m\rangle
\right\}_{m=-s}^{s}$, have the same kets, except that they are in
reverse order: the ket $|i\rangle$ of $B_d$ is equal to the ket
$|s-i\rangle$ of $B_s$.

The interest in some types of supersinglets is not new. For
instance, the two-particle $d$-level supersinglets $|{\cal
S}_2^{(d)}\rangle$ attracted some attention in connection with
violations of Bell's inequalities for systems of two spin-$s$
particles
\cite{Mermin80,MS82,GM82,GM83,Ogren83,BC88,SS90,Ardehali91,Peres92,GP92}.
These states can be expressed as
\begin{equation}
\left| {\cal S}_2^{(d)} \right\rangle = {1 \over \sqrt{d}}
\sum_{i=0}^{d-1} (-1)^i |i\rangle \otimes |d-i-1\rangle.
\label{s2sN}
\end{equation}
Examples of $| {\cal S}_2^{(d)} \rangle$ states are the following:
\begin{eqnarray}
\left| {\cal S}_2^{(2)} \right\rangle & = & {1 \over \sqrt{2}}
\left( |01\rangle - |10\rangle \right),
\label{s2s2}
\\
\left| {\cal S}_2^{(3)} \right\rangle & = & {1 \over \sqrt{3}}
\left( |02\rangle - |11\rangle + |20\rangle \right), \label{s2s3}
\\
\left| {\cal S}_2^{(4)} \right\rangle & = & {1 \over 2} \left(
|03\rangle - |12\rangle + |21\rangle - |30\rangle \right).
\label{s2s4}
\end{eqnarray}
The $|{\cal S}_2^{(3)}\rangle$ state has also been used in
two-particle proofs of Bell's theorem without inequalities
\cite{HR83,BS90,Cabello98}. A method for preparing optical analogs
of $|{\cal S}_2^{(d)}\rangle$ for every $d$ has been recently
described \cite{LHB01}. Also, an experimental violation of a
Bell's inequality using an optical analog of $|{\cal
S}_2^{(3)}\rangle$ has been recently reported \cite{HLB01}.

In this paper we are interested in two types of supersinglets:
$N$-particle $N$-level supersinglets, $|{\cal S}_N^{(N)}\rangle$,
and $N$-qubit (with $N$ even) supersinglets, $|{\cal
S}_N^{(2)}\rangle$. The former can be expressed as
\begin{equation}
\left|{\cal S}_N^{(N)} \right\rangle={1 \over \sqrt {N!}}
\sum_{\scriptscriptstyle{ {\stackrel{\scriptscriptstyle{\rm
permutations}} {{\rm of}\;01\ldots(N-1)}}}} \!\!\!\!\!\!
(-1)^t\left| ij\ldots n \right\rangle,
\label{SNSN}
\end{equation}
where $t$ is the number of transpositions of pairs of elements
that must be composed to place the elements in canonical order
(i.e., $0,1,2,\ldots,N-1$). The $|{\cal S}_2^{(2)}\rangle$ state
is given in equation~(\ref{s2s2}). Other examples of $|{\cal
S}_N^{(N)}\rangle$ states are the following:
\begin{eqnarray}
\left|{\cal S}_3^{(3)} \right\rangle & = & {1 \over \sqrt {6}} (
\left| 012 \right\rangle - \left| 021 \right\rangle -
\left| 102 \right\rangle + \left| 120 \right\rangle \nonumber \\
& & + \left| 201 \right\rangle - \left| 210 \right\rangle ),
\label{s3s3} \\
\left|{\cal S}_4^{(4)} \right\rangle & = & {1 \over \sqrt {24}} (
\left| 0123 \right\rangle - \left| 0132 \right\rangle - \left|
0213 \right\rangle + \left| 0231 \right\rangle \nonumber \\ & & +
\left| 0312 \right\rangle - \left| 0321 \right\rangle
- \left| 1023 \right\rangle + \left| 1032 \right\rangle \nonumber
\\ & & + \left| 1203 \right\rangle - \left| 1230 \right\rangle -
\left| 1302 \right\rangle + \left| 1320 \right\rangle
\nonumber \\ & & + \left| 2013 \right\rangle - \left| 2031
\right\rangle -
\left| 2103 \right\rangle + \left| 2130 \right\rangle \nonumber \\
& & + \left| 2301 \right\rangle - \left| 2310 \right\rangle
- \left|
3012 \right\rangle + \left| 3021 \right\rangle \nonumber \\
& & + \left| 3102 \right\rangle - \left| 3120 \right\rangle -
\left| 3201 \right\rangle + \left| 3210 \right\rangle).
\label{s4s4}
\end{eqnarray}

The $|{\cal S}_3^{(3)} \rangle$ state has been used in the
solution of the so-called {\em Byzantine agreement problem}
\cite{FGM01} (see more on this problem in section~\ref{sec:3}).
The $|{\cal S}_N^{(N)} \rangle$ states have been used in a scheme
designed to probe a quantum gate that realizes an unknown unitary
transformation \cite{HB01}. Some interesting properties of $|{\cal
S}_3^{(3)} \rangle$ have been reported in \cite{DW02}. An
experimental realization of $|{\cal S}_3^{(3)} \rangle$ has been
recently proposed \cite{Gisin02}.


For $N$ even, the $N$-qubit supersinglet can be expressed as
\begin{eqnarray}
\left|{\cal S}_N^{(2)} \right\rangle = {1 \over {N \over 2}! \sqrt
{{N \over 2}+1} }\sum_{\scriptscriptstyle{
{\stackrel{\scriptscriptstyle{\rm permutations}} {{\rm
of}\;0\ldots01\ldots1}}}} \!\!\!\!\!\! z! \left( {N \over 2}-z
\right)! (-1)^{{N \over 2}-z} \nonumber \\ \left| ij\ldots n
\right\rangle, \label{SnS2}
\end{eqnarray}
where the sum is extended to all the states obtained by permuting
the state $|0\ldots 01\ldots1\rangle$, which contains the same
number of zeros and ones; $z$ is the number of zeros in the first
$N / 2$ positions (for example: in $|01\rangle$, $z=1$; in
$|1100\rangle$, $z=0$; in $|010110\rangle$, $z=2$). The $|{\cal
S}_2^{(2)}\rangle$ state is given in equation (\ref{s2s2}). Other
examples of $|{\cal S}_N^{(2)}\rangle$ (with $N$ even) states are
the following:
\begin{eqnarray}
\left|{\cal S}_4^{(2)} \right\rangle & = & {1 \over 2 \sqrt{3}}
(2|0011\rangle-|0101\rangle-|0110\rangle \nonumber \\
& & -|1001\rangle-|1010\rangle+2|1100\rangle), \\
\left|{\cal S}_6^{(2)} \right\rangle & = & {1 \over 6}
(3|000111\rangle-|001011\rangle-|001101\rangle \nonumber \\
& & -|001110\rangle-|010011\rangle-|010101\rangle \nonumber \\
& & -|010110\rangle+|011001\rangle+|011010\rangle \nonumber \\
& & +|011100\rangle-|100011\rangle-|100101\rangle \nonumber \\
& & -|100110\rangle+|101001\rangle+|101010\rangle \nonumber \\
& & +|101100\rangle+|110001\rangle+|110010\rangle \nonumber \\
& & +|110100\rangle-3|111000\rangle).
\end{eqnarray}
Very recently, the $|{\cal S}_4^{(2)} \rangle$ state has been
experimentally prepared using four parametric down-converted
photons entangled in polarization \cite{Weinfurter02}. Some
applications of the $|{\cal S}_N^{(2)} \rangle$ states are
described in section~\ref{sec:5}.

The structure of this paper is the following: In
section~\ref{sec:2}, three apparently unrelated problems are
introduced. In section~\ref{sec:3}, these three problems are
proved to not have classical solution. In section~\ref{sec:4} a
solution based on the use of the $|{\cal S}_N^{(N)} \rangle$
states is presented. Finally, in section~\ref{sec:5}, the use of
the $|{\cal S}_N^{(2)} \rangle$ states to create decoherence-free
subspaces with qubits is described.


\section{Three problems}
\label{sec:2}


\subsection{The $N$ strangers problem}


The scenario for the {\em $N$ strangers problem} (NSP) is
an extension to a high number $N$ of players of the situation
described in Patricia Highsmith's novel \cite{Highsmith50} and
Alfred Hitchcock's movie \cite{Hitchcock51} {\em Strangers on a
Train}: $N$ complete strangers $A_i$ ($i=1,\ldots,N$) meet on a
train. $A_i$ wants $B_i$ dead. In the course of small talk, one suggests
that an ``exchange'' murder between $N$ complete strangers would
be unsolvable. After all, how could the police find the murderer
when he/she is a total and complete stranger with absolutely no
connection whatsoever to the victim? $A_i$ could kill
$B_k$, etc. However, such a plan suffers from an important
problem: if all the players know who the murderer of each victim
is, then the whole plan is vulnerable to individual denunciations.
Alternatively, if the distribution of victims is the result of a
secret lottery, how could the murderers be assured that the
lottery was not rigged and that nobody had contrived the result or
could ascertain it?

In more general, noncriminal, terms the $N$ strangers problem can
be reformulated as follows: Each of $N$ parties must perform one
among $N$ different tasks which they do not want to perform
themselves. However, each party volunteers to perform some other
party's task.

The problem is then how to distribute the victims/tasks
$\{B_i\}_1^{N}$ among the murderers/volunteers $\{A_i\}_1^{N}$,
which share no previous secret information nor any secure
classical channel, in a way that guarantees that each
murderer/volunteer $A_i$ knows only the identity of his/her
victim/task and that nobody else (besides the murderers/volunteers)
knows anything about the victim/task assignments.


\subsection{The secret sharing problem}


The {\em secret sharing problem} (SSP) was already described, for
$N=3$, in \cite{HBB99}. It could arise in the following context:
$A_1$ wants to have a secret action taken on her behalf at a
distant location. There she has $N-1$ agents, $A_2$, $A_3$,
\ldots, $A_N$ who carry it out for her. $A_1$ knows that some of
them are dishonest, but she does not know which ones. She
cannot simply send a secure message to all of them, because the
dishonest ones will try to sabotage the action, but it is assumed
(as in \cite{HBB99}) that if all of them carry it out together,
the honest ones will keep the dishonest ones from doing any
damage. The problem is then that $A_1$ wishes to convey a
cryptographic key to $A_2$, $A_3$, \ldots, $A_N$ in such a way
that none of them can read it on their own, only if all the $A_i$
($i=2,3,\ldots,N$) collaborate. In addition, they wish to prevent
any eavesdropper from acquiring information without being
detected. It is assumed that $A_1$ shares no previous secret
information nor any secure classical channel with her agents.
Different quantum solutions to this problem for $N=3$ has been
proposed using either GHZ \cite{ZZHW98,HBB99} or Bell states
\cite{KKI99} (see also \cite{Gruska99}). Below we shall propose a
different solution for any $N$.


\subsection{The liar detection problem}


The {\em liar detection problem} (LDP) occurs in the following
scenario: three parties $A$, $B$, and $C$ are connected by secure
pairwise classical channels. Let us suppose that $A$ sends a
message $m$ to $B$ and $C$, and $B$ sends the same message to $C$.
If both $A$ and $B$ are honest, then $C$ should receive the same
$m$ from $A$ and $B$. However, $A$ could be dishonest and send
different messages to $B$ and $C$, $m_{AB} \neq m_{AC}$, or,
alternatively, $B$ could be dishonest and send a message which is
different from the one he receives, $m_{BC} \neq m_{AB}$. For $C$
the problem is to ascertain without any doubt who is being
dishonest. This problem is interesting for classical information
distribution in pairwise connected networks. The message could be
a database and the dishonest behavior a consequence of an error
during the copying or distribution process.



\section{Proofs of impossibility of classical solutions}
\label{sec:3}


In this section I shall explain in which sense each of the three
problems introduced in section~\ref{sec:2} have no classical
solution. I will furthermore offer proofs of the impossibility of
the existence of that kind of solutions.

In Peres' book \cite{Peres93} (p.~293), the {\em key distribution
problem} (KDP) is defined as follows: ``The problem (\ldots) is
how to distribute a cryptographic key (a secret sequence of bits)
to several observers who initially share no secret information, by
using an insecure communication channel subject to inspection by a
hostile eavesdropper. If only classical means are used, this is an
impossible task. Quantum phenomena, on the other hand, provide
various solutions. The reason for this difference is that
information stored in classical form (\ldots) can be examined
objectively without altering it in any detectable way, let alone
destroying it, while it is impossible to do that with quantized
information encoded in unknown non-orthogonal states.''

In this definition, trusted couriers, clandestine meetings or
private secure communication links between the parties are not
allowed. It is in this scenario---namely, one in which previous
secret information and secure communication channels between all
parties (in the NSP) or between $A_1$ and her agents (in the SSP)
are not allowed, but public classical channels which cannot be
altered are assumed---in which the first two problems (NSP and
SSP) have no classical solution. A proof follows.

\begin{lemma}
The KDP between two parties has no classical solution.
\end{lemma}

{\em Proof:} Information stored in classical form can be examined
objectively without altering it in any detectable way, {\em
q.e.d.}\hfill\endproof

\begin{lemma}
The NSP has no classical solution.
\end{lemma}

{\em Proof:} Suppose the NSP had a classical solution for some
$N>1$. Then, $N-2$ parties could publicly announce the names of
their victims and then the remaining two parties would have a way
to solve the KDP (for instance, by publicly assigning 0 and 1 to
the remaining two victims and agreeing to define the corresponding
entry of the key as the bit value assigned to the victim of one of
the two parties), {\em q.e.d.}\hfill\endproof

\begin{lemma}
The SSP has no classical solution.
\end{lemma}

{\em Proof:} Suppose the SSP had a classical solution for some
$N>2$ (if $N=1$, then the SSP is the KDP). Such a solution must
work even when all of $A_1$'s agents are honest. In this case,
$N-2$ agents could publicly collaborate and share the result of
such a collaboration with the remaining agent. Then $A_1$ and the
remaining agent would have a way to solve the KDP, {\em
q.e.d.}\hfill\endproof\\

The scenario of the third problem, the LDP, is different. $A$,
$B$, and $C$ are now connected by pairwise classical channels. These
channels must be secure. In this scenario---namely, one in which
previous secret information is not allowed, but secure pairwise
classical channels are assumed---, the LDP has no classical
solution. In order to prove this, we need some previous results.

The {\em Byzantine generals problem} (BGP) is defined
\cite{PSL80,LSP82} as follows: A commanding general must send an
order to his $N-1$ lieutenant generals such that: (IC1) All loyal
lieutenant generals obey the same order. (IC2) If the commanding general
is loyal, then every loyal lieutenant general obeys the order he sends.
(IC1) and (IC2) are known as the interactive consistency
conditions. Note that if the commanding general is loyal, (IC1) follows from
(IC2). However, the commanding general may be a traitor.

\begin{lemma}
The BGP for $N=3$ generals and $1$ traitor has no classical
solution.
\end{lemma}

{\em Proof} \cite{LSP82} (p.~384): For simplicity, we consider the
case in which the only possible orders are ``attack'' or
``retreat''. Let us first examine the scenario in which the
commanding general is loyal and sends an ``attack'' order, but lieutenant general~2
is a traitor and reports to lieutenant general~1 that he received a
``retreat'' order. For (IC2) to be satisfied, lieutenant general~1 must
obey the order to attack. Now consider another scenario in which
the commanding general is a traitor and sends an ``attack'' order to
lieutenant general~1 and a ``retreat'' order to lieutenant general~2. Lieutenant general~1
does not know who the traitor is, and he cannot tell what message
the commanding general actually sent to lieutenant general~2. Hence, the scenarios
in these two cases appear exactly the same to lieutenant general~1. If the
traitor lies consistently, then there is no way for lieutenant general~1
to distinguish between these two situations, so he must obey the
``attack'' order in both of them. Hence, whenever lieutenant general~1
receives an ``attack'' order from the commanding general, he must obey it,
{\em q.e.d.}\hfill\endproof\\

For a more rigorous proof, see \cite{PSL80}.

\begin{lemma}
No solution for the BGP for $N <
3M+1$ generals can cope with $M$ traitors.
\end{lemma}

{\em Proof:} The BGP for $N < 3M+1$ generals and $M$ traitors can
be reduced to the BGP for $N=3$ generals and $M=1$ traitor, with
each of the generals simulating at most $M$ lieutenants and taking
the same decision as the loyal lieutenants they simulate, {\em
q.e.d.}\hfill\endproof\\

A solution for the BGP for $N > 3M$ and up to $M$ traitors is
given in \cite{LSP82}.

The protocol proposed in \cite{FGM01} solves the BGP for $N=3$ in
the following sense: (1) If all generals are loyal, then the
protocol achieves broadcast. (2) If one general is a traitor, then
either the protocol achieves broadcast or both loyal generals
abort the protocol.

The LDP described in section~\ref{sec:2} is simplification of the
BGP for $N=3$ and $M=1$.

\begin{lemma}
The LDP has no classical solution.
\end{lemma}

{\em Proof:} Supposing that the LDP had a classical solution, then
the BGP for $N=3$ and $M=1$ had classical solution, {\em
q.e.d.}\hfill\endproof


\section{Solutions}
\label{sec:4}


In this section I will show that all three problems introduced in
section~\ref{sec:2} can be solved if each of the $N$ participants
are in possession of a sequence of numbers with the following
properties:
\begin{itemize}
\item[(i)] It is {\em random} (i.e.,
generated by an intrinsically unrepeatable method which gives each
possible number with the same probability of occurring).
\item[(ii)] The possible numbers are integers from $0$ to $N-1$.
\item[(iii)] If a number $i$ is at position $j$ of the sequence of
party $k$, $i$ is not at position $j$ in the sequence of a
different party.
\item[(iv)] Each party knows only his/her own
sequence.
\item[(v)] Nobody else (besides the parties) knows the
sequences. \end{itemize}
Properties (iv) and (v) are difficult to
accomplish using classical tools due to the fact that information
transmitted in classical form can be examined and copied without
altering it in any detectable way. However, as quantum key
distribution protocols show \cite{BB84,Ekert91}, quantum
information does not suffer from such a drawback. Assuming we have
a reliable method of generating sequences of numbers with properties
(i) to (v) among $N$ distant parties, a method that will be
presented below, then the solutions to the above problems are
described in the following subsections.


\subsection{Solution to the $N$ strangers problem}


Each victim $B_i$ is assigned a label, taken from $0$ to $N-1$. If
murderer $A_i$'s sequence starts with $j$, then $A_i$ must kill
$B_j$, etc. The remaining entries of the sequence can be used for
subsequent rounds. The result tells every murderer who his/her
victim is in such a way that prevents any murderer (or even a
small group of them) from denouncing or blackmailing another. The
only way to ascertain with certainty who murdered $B_j$ is that
all the other murderers confess. $A_i$ has a probability of $1/N$
to have to kill his/her own desired victim $B_i$. In case $A_i$
must kill $B_i$, since the fact that $A_i$ wants $B_i$ dead is
presumably known by others, the best thing $A_i$ could do is to
participate in another round with a different set of strangers.


\subsection{Solution to the secret sharing problem}


The key is defined as $A_1$'s sequence. The only way to reveal it
is to make the remaining $N-1$ parties share their respective
sequences; the key is then composed by the missing results. If a
dishonest party $D$ declares a result which is different to
his/her actual result, then there is a probability $1/(r-1)$,
where $r$ is the number of honest parties which have not yet
declared their results, that other honest party $H$ has obtained
that result. Then $H$ would stop the process, so Alice's key (and
thus Alice's action) would remain safe (dishonest parties cannot
sabotage Alice's action if they do not know what it is). The order
in which the agents declare their respective results must change
from round to round to avoid any dishonest party being always the
last to declare.


\subsection{Solution to the liar detection problem}


Before entering into the details of the solution, it will be
useful to sketch how it works. $A$(lice) must send $B$(ob) [and
$C$(arol)] some information $l_{AB}{(m_{AB})}$
[$l_{AC}{(m_{AC})}$] such that: (a) It is only known to Alice. (b)
Its authenticity can be checked by $B$ (and $C$) using information
only known to him (her). (c) The information $l_{AB}{(m_{AB})}$
[$l_{AC}{(m_{AC})}$] is correlated with the message $m_{AB}$
($m_{AC}$); if the message were different, then this
information would also be different. Bob needs $l_{AB}{(m_{AB})}$ to
convince Carol that the message $A$ sent him was actually $m_{AB}$
(and not other).


\begin{figure}
\centerline{\includegraphics[width=8.2cm]{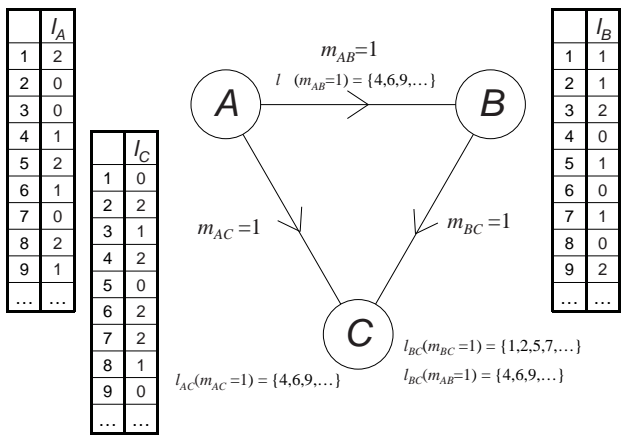}}
\caption{\label{liard} Protocol for solving the liar detection
problem. To explain how it works, we will assume that nobody is
lying. $A$ must send $B$ a message $m_{AB}$ and the list
$l_{AB}{(m_{AB})}$ of positions in $A$'s list $l_A$ (left) in
which the message $m_{AB}$ appears. Analogously, $A$ must send $C$
a message $m_{AC}$ and the corresponding list $l_{AC}{(m_{AC})}$.
Finally, $B$ must send $C$ a message $m_{BC}$ and two lists:
$l_{BC}{(m_{BC})}$ and $l_{AB}{(m_{AB})}$.}
\end{figure}


Let us suppose that the message $m$ is a trit value $0$, $1$, or
$2$. All three parties agree to use the following protocol:
\begin{itemize}
\item[(I)] If the transmitted message is $m_{ij}$, then the sender
$i$ must also send $j$ the list $l_{ij}{(m_{ij})}$ of positions in
his/her sequence in which the number $m_{ij}$ appears (see
figure~\ref{liard}). Note that if the sequences are random and
long enough then any $l_{ij}{(m_{ij})}$ must contain about one
third of the total length $L$ of the sequences.
\item[(II)] The receiver $j$ would
not accept any message if the intersection between the received
list $l_{ij}{(m_{ij})}$ and his/her list $l_{j}{(m_{ij})}$ is not
null nor if $l_{ij}{(m_{ij})} \ll L/3$ elements.
\end{itemize}
We will assume that requirements (I) and (II) force the dishonest
party to send correct but perhaps incomplete lists. Otherwise, if
$i$ sends a list containing $n$ erroneous data, then the
probability that $j$ does not accept the message $m_{ij}$ would be
$(2^n-1)/2^n$. In addition,
\begin{itemize}
\item[(III)] $B$ must send $C$ the list $l_{BC}{(m_{AB})}$ containing
the sequence he has (supposedly) received from $A$ (see
figure~\ref{liard}).
\end{itemize}
Therefore, when $C$ finds that $m_{AC} \neq m_{BC}$, she has
received three lists to help her to find out whether it is $A$ or
$B$ who is being dishonest (see figure~\ref{liard}). According to
rules (I) to (III), if $B$ wants to be dishonest $l_{BC}{(m_{BC})}
\cup l_{BC}{(m_{AB})}$ must necessarily be a subset of
$l_{B}{(m_{BC})}$, because $B$ does not know $l_A{(m_{BC})}$.
However, the length of $l_{B}{(m_{BC})}$ is about $L/3$, while $C$
is expecting $B$ to send her two lists with a total length of $2
L/3$; then $C$ would conclude that $B$ was being dishonest.
Alternatively, if it is $A$ who is being dishonest, the length of
the two lists that $C$ received from $B$ would total about $2
L/3$; $C$ would then conclude that $A$ was being dishonest.


\subsection{$N$-particle $N$-level supersinglets}


A possible quantum method to generate
sequences of numbers with properties (i) to (iv) among $N$ distant parties is by
distributing among all $N$ parties an $N$-particle $N$-level
supersinglet $|{\cal S}_N^{(N)}\rangle$ described in~(\ref{SNSN}).
Once the particles have been distributed among the parties, a
direction of measurement is randomly chosen and publicly
announced. Due to the $N$-lateral rotational invariance of the
$|{\cal S}_N^{(N)}\rangle$ states, whenever the $N$ parties
measure the spin of the $N$ separated particles along this
direction, each of them finds a different result in the set
$\{0,\ldots,N-1\}$; thus such results satisfy requirements~(ii)
and~(iii).

In order to accomplish requirements (i), (iv), and (v), an
essential property is {\em nonseparability}, that is, the quantum
predictions for the $|{\cal S}_N^{(N)}\rangle$ states cannot be
reproduced by any local hidden variables model in which the
results of the spin measurements are somehow determined before the
measurement. To show the nonseparability of $|{\cal S}_N^{(N)}\rangle$
we have to study whether they violate Bell's inequalities derived
from the assumptions of local realism. Most Bell's inequalities
require two alternative local dichotomic (taking values $-1$
or $1$) observables $A_j$ and $B_j$ on each particle $j$. To test
nonseparability, we will use the dichotomic local observables
proposed by Peres in \cite{Peres92}. A Peres' observable $A_k$ can
be operationally defined as follows: to measure $A_k$, first
measure the spin component of particle $k$ along direction $A$,
$S_A^{(k)}$. If particle $k$ is a spin-$s$ particle, then
measuring $S_A^{(k)}$ could give $2 s+1$ different results. Then
assign value $1$ to results $s$, $s-2$, etc., and value $-1$ to
results $s-1$, $s-3$, etc. The operator representing observable
$A_k$ can be written as
\begin{equation} \hat A_k =
\sum_{m=-s}^{s} (-1)^{s-m} |S_A^{(k)}=m \rangle \langle
S_A^{(k)}=m |, \label{Peresob}
\end{equation} where
$|S_A^{(k)}=m\rangle$ is the eigenstate of the spin component
along direction $A$ of particle $k$.

To show the nonseparability of the $|{\cal S}_N^{(N)}\rangle$ states,
let us consider the following scenario: $N$ distant observers
share $N$ $N$-level particles in the $|{\cal S}_N^{(N)}\rangle$ state;
the $N-m$
observers can choose between measuring $A_j=A$ and $B_j=a$; the
remaining $m$ observers can choose between measuring $A_k=B$ and
$B_k=b$. Then nonseparability can be tested by means of the
following Bell's inequality, which generalizes to $N$ particles
the Clauser-Horne-Shimony-Holt inequality \cite{CHSH69}
\begin{eqnarray}
|E_N(A,\ldots,A,B, \ldots,B)+ E_N(A,\ldots,A,b,\ldots,b) \nonumber \\
+E_N(a,\ldots,a,B,\ldots,B)-E_N(a,\ldots,a,b,\ldots,b)| \le 2.
\label{CHSHg}
\end{eqnarray}
Note that this inequality uses only a subset of
all possible correlation functions (for instance, it does not use
$E_N(A,a,\ldots,a,B,\ldots,B)$). For the $|{\cal S}_N^{(N)}\rangle$ states,
restricting our attention to Peres' observables, the
correlation function $E_N^{(m)}(A,\ldots,A,B,\ldots,B)$, which
represents the expected value of this product of the results of
measuring, for instance, $N-m$ observables $A$, and $m$
observables $B$ is given by
\begin{eqnarray}
\!\!E_N^{(N-1)} &\!\!
= &\!\! (-1)^{f(N/2)} {1 \over N}
{\sin (N \theta_{AB}) \over \sin \theta_{AB}}, \\
E_N^{(N-2)} &\!\! = &\!\! (-1)^{f(N/2)} {1 \over N+2}
\left\{1+{\sin [(N+1) \theta_{AB}] \over \sin
\theta_{AB}}\right\},
\end{eqnarray}
where $\theta_{AB}$ is the
angle between directions $A$ and $B$ and $f(x)$ gives the greatest
integer less than or equal to $x$. In case of $m=1$, that is,
using correlation functions of the $E_N^{(N-1)}$ type, we have
found that $|{\cal S}_N^{(N)}\rangle$ states violate inequality
(\ref{CHSHg}) for any $N$. The maximum violation for $N=2$ is $2
\sqrt{2}$, for $N=3$ is $2.552$, and for $N \rightarrow \infty$
tends to $2.481$. In case of $m=2$, that is, using correlation
functions of the $E_N^{(N-2)}$ type, we have found that the
$|{\cal S}_N^{(N)}\rangle$ states violate inequality (\ref{CHSHg}) for any $N$.
The maximum violation for $N=4$ is $2.418$, for $N=5$ is $2.424$
and for $N \rightarrow \infty$ tends to $2.481$.

So far we have assumed that the $N$ parties share a large
collection of $N$-level systems in the $|{\cal S}_N^{(N)}\rangle$ state.
To make sure that this is the case (and that nobody has changed the
state) requires a protocol to distribute and test these states
between the $N$ parties such that at the end of the protocol
either all parties agree that they share a $|{\cal S}_N^{(N)}\rangle$
state (and then they can reliably apply the described solutions),
or all of them conclude that something went wrong (and then abort
any subsequent action). For $N=3$ such a distribute-and-test
protocol is explicitly described in \cite{FGM01} and can be easily
generalized to any $N>3$. The test requires that the parties
compare a sufficiently large subset of their particles which are
subsequently discarded.


\section{$N$-qubit supersinglets and decoherence-free subspaces}
\label{sec:5}


$N$-qubit (with $N$ even) supersinglets $|{\cal S}_N^{(2)}\rangle$
have found many applications: they can be used to solve the KDP,
and the SSP, defined in section~\ref{sec:2}; they can be used to
do {\em telecloning} (a process combining quantum teleportation
and optimal quantum cloning from one input to $M$ outputs
\cite{MJPV99}). In fact, $|{\cal S}_N^{(2)}\rangle$ are the
$N$-lateral rotationally invariant version of the telecloning
states introduced in \cite{MJPV99}. Recently, it has been found
that $|{\cal S}_4^{(2)}\rangle$ can also be used to solve the LDP
described in section~\ref{sec:2} \cite{Cabello02}. Here, however,
we shall be examining a different use for these states: encoding
quantum information in a way that makes sure that it is not
affected by some kind of otherwise unavoidable errors.

In quantum computation and quantum communication \cite{NC00},
information is stored in the quantum state of quantum systems,
usually qubits. Preserving the desired states of the qubits and
controlling their evolution is essential for most of the tasks.
However, when the qubits interact with the environment, although
the overall evolution of the state of the qubits-environment
system is unitary (according to quantum mechanics, the evolution
of the state of any closed system is unitary), the evolution of the
state of the qubits alone (which do not form a closed system) is
not unitary. The state of the qubits rapidly evolves into a
state which is entangled with the environment. This implies that, even if
the initial state of the qubits was pure ($\rho^2=\rho$), after
such an interaction the state of the qubits rapidly becomes mixed
($\rho^2\neq\rho$). This process is know as {\em decoherence}
\cite{Zurek81,Zurek82}. Decoherence is thus a major obstacle in
quantum computation and quantum communication.

There are two ways to overcome decoherence. One is based on adding
redundancy when encoding information in order to detect and correct
errors by using {\em quantum error correction codes}
\cite{Shor95,Steane96,EM96,CS96,Gottesman96,CRSS97}. A quantum
error-correcting code is defined as a unitary mapping
(encoding) of $k$ qubits into a subspace of the quantum state
space of $n$ qubits such that, if any $t$ of the qubits undergos
arbitrary decoherence, not necessarily independently, the
resulting $n$ qubits can be used to faithfully reconstruct the
original quantum state of the $k$ encoded qubits.

The other approach is by encoding qubits within subspaces which do
not suffer decoherence due to reasons of symmetry. These
subspaces are called {\em decoherence-free (DF) subspaces}
\cite{PSE96,DG97,ZR97a,ZR97b,Zanardi97,DG98a,Zanardi98,DG98b,LCW98,LBW99,LBKW00,LBKW01a,LBKW01b}.
Indeed, DF subspaces can be considered as a special class of
quantum error correction codes \cite{LBW99}.

Let us illustrate the use of the $|{\cal S}_N^{(2)}\rangle$ states
for constructing DF subspaces by considering a simple example: $N$
qubits coupled by a single thermal bath described by a collection
of noninteracting linear oscillators. Let us assume that all
qubits suffer the {\em same} interaction with the bath. Such an
assumption is justified as long as the qubits have very close
positions with respect to the bath coherence length. Then, it has
been shown that, for $N$ even, the $|{\cal S}_N^{(2)}\rangle$
states are eigenstates of the whole Hamiltonian of the qubits-bath
system and also eigenstates of the interaction Hamiltonian with
eigenvalue zero (see \cite{ZR97a} for details). This is due to the
fact that the $|{\cal S}_N^{(2)}\rangle$ states are $N$-lateral
unitary invariant. This means that, if we act on any of them with
the tensor product of $N$ equal unitary operators, the result will
be to reproduce the same state:
\begin{equation}
U^{\bigotimes N} \left|{\cal S}_N^{(2)}\right\rangle= \left|{\cal
S}_N^{(2)}\right\rangle,
\end{equation}
$U^{\bigotimes N}$ being $U \otimes \ldots \otimes U$, where $U$
is an unitary operator \cite{comm}.

In addition, any (coherent or incoherent)
superposition of $N$-lateral unitary invariant states
is also $N$-lateral unitary invariant. Therefore, for arbitrary $N$
(even), the $|{\cal S}_j^{(2)} \rangle$ states with $j$ ranging
from 2 to $N$ (and tensor products thereof) span a
$d(N)$-dimensional DF subspace. Let us call this subspace ${\cal C}_{d(N)}$.
The dimension of ${\cal C}_{d(N)}$ is given \cite{ZR97a} by
\begin{equation}
d(N)={N! \over (N/2)! (N/2+1)!}.
\end{equation}
The number of qubits encoded in ${\cal C}_{d(N)}$ is $\log_2 d(N)$.
For large $N$,
\begin{equation}
\log_2 d(N) \simeq N-{3 \over 2} \log_2 N.
\end{equation}
Therefore, the encoding efficiency, defined as
\begin{equation}
{\cal E}(N) \equiv {\log_2 d(N) \over N},
\end{equation}
is asymptotically {\em unity}.


\begin{acknowledgments}
I thank N. D. Mermin and C. Serra for useful comments, M.
Bourennane, M. Eibl, S. Gaertner, N. Kiesel, and H. Weinfurter for
illuminating discussions, M. Ferrero and the rest of organizers
for inviting me to the Oviedo Conference on Quantum Information,
H. Weinfurter for his hospitality at
Ludwig-Maximilians-Universit\"{a}t, M\"{u}nchen, and
Max-Planck-Institut f\"{u}r Quantenoptik, Garching, and the
Spanish Ministerio de Ciencia y Tecnolog\'{\i}a grants
BFM2001-3943 and BFM2002-02815, the Junta de Andaluc\'{\i}a grant
FQM-239, and the Max-Planck-Institut f\"{u}r Quantenoptik for
support.
\end{acknowledgments}




\begin{thebibliography}{99}
\bibitem{comm}
The $|{\cal S}_N^{(N)}\rangle$ states given in (\ref{SNSN}) are
also $N$-lateral unitary invariant, but not the $|{\cal
S}_2^{(d)}\rangle$ states given in (\ref{s2sN}), which are simply
$N$-lateral rotationally invariant.
\end{thebibliography}
\end{document}